\documentclass{article}
\usepackage{graphicx}

 \newcommand{\be}{\begin{equation}}
 \newcommand{\ee}{\end{equation}}
 \newcommand{\ben}{\begin{eqnarray}}
 \newcommand{\een}{\end{eqnarray}}

  \usepackage[dvipdf]{epsfig}
\usepackage{color}

\begin{document}

%
%

\title{Cosmological model with fermion and tachyon fields  interacting via Yukawa-type potential}

\author{Marlos O. Ribas\footnote{gravitam@yahoo.com.br}\\
Departamento de F\'{\i}sica, Universidade Tecnol\'ogica Federal do Paran\'a\\
 Curitiba, Brazil\\
Fernando P. Devecchi\footnote{devecchi@fisica.ufpr.br} and Gilberto M. Kremer\footnote{kremer@fisica.ufpr.br}\\
Departamento de F\'{\i}sica,
Universidade Federal do Paran\'a\\ Curitiba, Brazil}

\maketitle


\begin{abstract}
A model for the universe with  tachyonic and  fermionic fields interacting through a Yukawa-type potential  is investigated.
It is shown that the tachyonic field
answers for the initial accelerated regime and for the subsequent decelerated regime so that it behaves as an inflaton at early times and as a matter field at intermediate times, while the fermionic field has the role of a dark energy constituent, since it leads to an accelerated regime at later times. The interaction between the fields via a Yukawa-type potential  controls the duration of the decelerated era, since  a stronger coupling makes a shorter decelerated period.

\end{abstract}


\section{Introduction}

Fermionic sources can be  responsible for  accelerated
periods in cosmological models \cite{4,1a,1b,1c,1d,2,3a,3b,3c,3d,3e,3f,4a}. In
the majority of these  models the fermionic field plays the role of the inflaton in the
early period of the universe and the role of dark energy for the old
universe. These fermionic sources has been
investigated using several
 approaches, with results including numerical solutions, exact solutions,
anisotropy-to-isotropy scenarios and cyclic cosmologies (see, for
example \cite{1a,1b,1c,1d,4,3a,3b,3c,3d,3e,3f}).

It is possible to combine these fermionic sources with other constituents, like canonical and non-canonical scalar fields. In particular, the inflaton and the dark energy  can be modeled in several ways. One sub-class of these models consider that constituent to be a  tachyon field, an idea which can be traced back to string theory models.    These  tachyonic models has been confronted with observational data,  and when compared to the usual canonical scalar field the physics arising is in general richer, due to non-linear effects \cite{Moschella}.

In general, depending on the initial conditions and on the interaction with other sources, these tachyons could promote accelerated periods in old universes, as a dark energy contribution \cite{C0,C,C1,C2,C3,C4,C5,C6}.  They can be consistent also on early  universes, where they could be associated to an inflationary period \cite{D0,D,D1,D2,D3,D4,D5,D6,D7}. Furthermore, in the papers \cite{E1,E2} the tachyon field was considered to be the responsible for both the inflationary period and the currently accelerated expansion. These effects do not follow as ad hoc imposed conditions, they are in fact a consequence of the dynamics of the model itself. The exotic nature of these  constituents do not contradict observational data, and the comparison to the canonical scalar field is an important point of the discussion.

The aim of this work is to analyze a model for the universe with two constituents,  a  tachyonic and a fermionic fields  interacting through a Yukawa-type potential  \cite{Ryder1,Ryder2}. The numerical solutions of the field equations show that the model can answer for the accelerated-decelerated-accelerated periods of the universe, with the tachyonic field playing the role of the inflaton at early times and as matter field at intermediate times, while the fermionic field can be considered as a dark energy field for later times, since it answers for the late accelerated period.

The work is structured as follows: in Section 2 we present the model through an action which is a sum of the Lagrangian densities of the gravitational, tachyonic, fermionic and Yukawa interaction fields. Dirac, Einstein and Klein-Gordon type field equations are obtained from the variation of the action with respect to the fields.
 The field equations for a homogeneous and isotropic spatially flat universe is presented in Section 3. In Section 4 we show a numerical solution of the field equations with the
accelerated-decelerated-accelerated  transition and  the evolution of the total
pressure and energy density of the sources of the gravitational field. The main
conclusion of the work are stated  Section 5. The metric signature used is
$(+,-,-,-)$ and units have been chosen so that $8\pi G=c=\hbar=1$.

\section{The model}

In the special  relativity  context the fundamental fields of nature are irreducible  representations  of the Lorentz group and among them we find the spinorial representations.
When we want to include gravitational effects we follow the principle of general covariance, switching from the Lorentz group to the diffeomorphisms. The problem here is that we do not have a spinorial representation of this group, and the solution is to use the tetrad formalism\cite{Wein,Wald}. The basic idea consists to associate to each point of space-time a set of coordinates
 $\xi^a$ that are locally inertial, where the metric is Minkowskian $\eta_{ab}$. This set connects the metric tensor through
\begin{equation}g_{\mu\nu}=e^a_\mu e^b_\nu\eta_{ab},
\qquad \hbox{where}\qquad e^a_\mu=\frac{\partial \xi^a}{\partial x^\mu},
\end{equation}
are  the tetrad or vierbein. In these expressions the Latin indexes correspond to the Minkowskian local space-time and the Greek indexes to the curved manifold. In this context, the derivatives that act upon the spinor field $\psi$ are replaced by their covariant counterparts by
\begin{equation}\partial_\mu\psi\rightarrow D_\mu\psi=\partial_\mu\psi-\Omega_\mu\psi, \qquad \partial_\mu\bar\psi\rightarrow D_\mu\bar\psi=\partial_\mu\bar\psi+\bar\psi\Omega_\mu.
\end{equation}
Here, $\bar\psi=\psi^\dagger\gamma^0$ denotes the spinor adjoint and the symbols  $\Omega_\mu$ are spin connections given by
\begin{equation}\Omega_\mu=-\frac{1}{4}g_{\sigma\rho}\left[\Gamma^\rho_{\mu\delta}-e^\rho_b
\left(\partial_\mu e^b_\delta\right)\right]\Gamma^\delta\Gamma^\sigma,
\end{equation}
where $\Gamma^\rho_{\mu\delta}$ are the Christoffel symbols. In the above equation the
Dirac-Pauli $\gamma^a$ are replaced by new matrices, according to the general covariance principle\cite{Wald}. They follow a generalized Clifford algebra $\{\Gamma^\mu,\Gamma^\nu\}=2g^{\mu\nu}$, where $\Gamma^\mu=e^\mu_a\gamma^a$.

Our next step is to establish the field equations for the model, which will follow from the  action
\begin{equation}\label{1}
S=\int\sqrt{-g}\left[\frac{1}{2}R+{\cal{L}}_D+{\cal{L}}_{T}+{\cal{L}}_{Y}\right]d^4x,
\end{equation}
where $R$ is the curvature scalar.

The dynamics of the   spinor field $\psi$ is encoded in the Lagrangian density \begin{equation}{\cal{L}}_D=\frac{i}{2}\left[\bar\psi\Gamma^\mu D_\mu\psi-\left(D_\mu\bar\psi\right)\Gamma^\mu\psi\right]-V(\bar\psi\psi),\end{equation}
with   $V(\bar\psi\psi)$ denoting the self-interacting potential of the spinor field, which is supposed to be a function of the bilinear $\bar\psi\psi$.

The tachyon field  $\phi$ is described by a Lagrangian density
\begin{equation}{\cal{L}}_T=-V(\phi)\sqrt{1-\partial_\mu\phi\partial^\mu\phi},
\end{equation}
where $V(\phi)$ is the self-interacting potential of the tachyon field.

We assume that the tachyon and the fermion fields interact through a Yukawa-type potential \cite{Ryder1,Ryder2}, whose contribution to the total Lagrangian is given by
\begin{equation}{\cal{L}}_Y=-\lambda \bar\psi\phi\psi.
\end{equation}

From the variations of the  action (\ref{1}) with respect to the spinor field and its adjoint it follows the Dirac equations
\ben\label{2}
i\Gamma^\mu D_\mu\psi-\frac{\partial V(\bar\psi\psi)}{\partial\bar\psi}-\lambda\phi\psi=0,
\qquad
iD_\mu\bar\psi\Gamma^\mu+\frac{\partial V(\bar\psi\psi)}{\partial\psi}+\lambda\bar\psi\phi=0.
\een

Furthermore, the variation of the action (\ref{1}) with respect to  $\phi$ gives an equation of Klein-Gordon type for the tachyon field which reads
\begin{equation}\label{3}
\nabla^\mu\nabla_\mu\phi+\frac{1}{2}\frac{\nabla_\mu\phi\nabla^\mu
\left(\nabla_\nu\phi\nabla^\nu\phi\right)}{\left(1-\nabla^\nu\phi\nabla_\nu\phi\right)}
+\frac{1}{V(\phi)}\frac{dV(\phi)}{d\phi}+\frac{\lambda}{V(\phi)}\bar\psi\psi
\sqrt{1-\nabla^\mu\phi\nabla_\mu\phi}=0.
\end{equation}

Einstein filed  equations are obtained from the variation of the action (\ref{1}) with respect to the tetrads
\begin{equation}\label{4}
R_{\mu\nu}-\frac{1}{2}g_{\mu\nu}R=-T_{\mu\nu}.
\end{equation}
Above $T_{\mu\nu}$ is the energy-momentum tensor, which has  the following expression
\ben\nonumber
T_{\mu\nu}=\frac{i}{4}\left[\bar\psi\Gamma^\mu D^\nu\psi+\bar\psi\Gamma^\nu D^\mu\psi-D^\nu\bar\psi\Gamma^\mu\psi-D^\mu\bar\psi\Gamma^\nu\psi\right]\\\label{4a}
+\frac{V(\phi)}{\sqrt{1-\nabla_\rho\phi\nabla^\rho\phi}}\nabla_\mu\phi\nabla_\nu\phi-g_{\mu\nu}\left({\cal{L}}_D+{\cal{L}}_{T}+{\cal{L}}_{Y}\right).
\een

\section{Field equations in the Robertson-Walker metric}

 In this section we shall determine the field equations for a homogeneous and isotropic spatially flat Universe  described by the Robertson-Walker metric
\begin{equation}ds^2=dt^2-a(t)^2\left(dx^2+dy^2+dz^2\right),\end{equation}
where  $a(t)$ denotes the cosmic scale factor. In this metric the tetrad components are given by $e^\mu_0=\delta^\mu_0$ and  $e^\mu_i=\delta^\mu_i/{a(t)}$, so that the Dirac matrices and the spin connections become
\ben
\Gamma^0=\gamma^0,\qquad
\Gamma^i=\frac{1}{a(t)}\gamma^i, \qquad\Omega_0=0, \qquad \Omega_i=\frac{1}{2}\dot a(t)\gamma^i\gamma^0.
\een
Here the dot refers to time derivative.

According to the hypothesis of homogeneity and isotropy the fermion and tachyon fields are exclusively functions of time and in this case the
 (\ref{2}) and  (\ref{3}) turn into
\ben\label{5a}
\dot\psi+\frac{3}{2}\frac{\dot a}{a}+i\gamma^0\frac{\partial V(\bar\psi\psi)}{\partial\bar\psi}+i\lambda\gamma^0\psi\phi=0,
\\\label{5b}
\dot{\bar\psi}+\frac{3}{2}\frac{\dot a}{a}-i\frac{\partial V(\bar\psi\psi)}{\partial\psi}\gamma^0-i\lambda\bar\psi\gamma^0\phi=0,
\\\label{5c}
\frac{\ddot\phi}{\left(1-\dot\phi^2\right)}+3\frac{\dot a}{a}\dot\phi+\frac{1}{V(\phi)}\frac{dV(\phi)}{d\phi}+\frac{\lambda}{V(\phi)}
\bar\psi\psi\sqrt{1-\dot\phi^2}=0.
\een

From the Einstein field equations (\ref{4}) together with the expression for the energy-momentum tensor (\ref{4a}) we obtain the Friedmann and acceleration equations
\begin{equation}\label{6}
\left(\frac{\dot a}{a}\right)^2=\frac{1}{3}\rho, \qquad \frac{\ddot a}{a}=-\frac{1}{6}(\rho+3p),
\end{equation}
where total energy density $\rho$ and the total pressure $p$ are given by
\ben\label{7a}
\rho=\frac{V(\phi)}{\sqrt{1-\dot\phi^2}}+\lambda\phi\bar\psi\psi+V(\bar\psi\psi),
\\\label{7b}
p=-V(\bar\psi\psi)+\frac{\partial V(\bar\psi\psi)}{(\partial\psi)}\frac{\psi}{2}+\frac{\bar\psi}{2}\frac{\partial V(\bar\psi\psi)}{(\partial\bar\psi)}-V(\phi)\sqrt{1-\dot\phi^2}.
\een

\section{Cosmological solutions}

In order to analyze cosmological solutions from the proposed model we have first to specify the self-interaction potentials of the fermion and tachyon fields.
The self-interaction potential   proposed for the fermion field is  $V(\bar\psi\psi)=\Lambda (\bar\psi\psi)^n$, with $n$ and $\Lambda$ being constants. This potential allows us to manipulate  the Dirac equations (\ref{5a}) and (\ref{5b}) in order to obtain the solution for the bilinear $\bar\psi\psi=C/a^3$,  where $C$ is an integration constant. For the tachyon field  the  self-interacting potential chosen is $V(\phi)=\xi/\phi^2$, where  $\xi$ is a constant. These expressions can be  combined with (\ref{7a}) and  (\ref{7b}) so that the energy density and the pressure become
\ben\label{8a}
\rho=\frac{\xi}{\phi^2\sqrt{1-\dot\phi^2}}+\frac{\lambda C\phi}{a^3}+\Lambda\left(\frac{C}{a^3}\right)^n
\\\label{8b}
p=(n-1)\Lambda \left(\frac{C}{a^3}\right)^n-\frac{\xi}{\phi^2}\sqrt{1-\dot\phi^2}
\een

To solve numerically the system of field equations composed by the time evolution of the tachyon field (\ref{5c}) and of the acceleration equation (\ref{6})$_2$  together with the expressions (\ref{8a}) and (\ref{8b}) for the energy density and pressure, we must specify the following constants : $\lambda$  that controls the intensity of the Yukawa coupling, $C$ which is associated with the initial condition of the bilinear $\bar\psi\psi$, $\Lambda$ and $\xi$ related to the intensity  of the self-interaction potential and the scalar potential, and  the exponent of the fermion self-interacting potential $n$.
 In our numerical simulations we have adopted:
$\lambda=10,$ $\xi=0.8,$ $\Lambda=1$, $C=0.0001$ and $n=0.5.$

Apart from the constants we have to specify initial conditions. Here we consider that initial values of the cosmic scale factor of the tachyon field and of the energy density are:
$a(0)=1$, $\phi(0)=1,$ and $\rho(0)=1$. With these values one can obtain from the Friedmann equation (\ref{6})$_1$ that $\dot a(0)=1/\sqrt3$ and from (\ref{7a}) that
\ben
\dot\phi(0)=\sqrt{1-\frac{\xi^2}{\phi(0)^4}\left(\frac{1}{1-\Lambda\left(\frac{C}{a(0)^3}\right)^n
-\frac{\lambda C\phi(0)}{a(0)^3}}\right)^2}.
\een

In figure  1 we show the behavior of the acceleration field as function of time, where we verify that for initial times there is an accelerated regime, followed  by a decelerated  and a final accelerated regimes. If the fermionic field is not present we have verified that there exist only an accelerated regime followed by a decelerated one, so that for later times the universe does not come back to an accelerated regime. These results permit us to conclude that the tachyon is playing the role of an inflaton for early times  and as non-relativistic matter for later times, while the fermionic field answers for a later and final  accelerated regime of the universe, behaving as a dark energy source.

As mentioned  above the coupling between the constituents occurs via
a Yukawa-type  potential.  What we verify is that this coupling  has an
important role on the extension in time of the decelerated regime: the
stronger the coupling (controlled by the $\lambda$  coefficient) the
earlier the universe is dominated by the dark energy constituent.  In fact
this a similar result to the one found in one of our previous
works\cite{3e}, where the universe was filled with  fermionic and  canonical
scalar fields: the Yukawa term was responsible for the extension in time
and the amplitude of the decelerated era.

The energy density and the pressure fields  decay with time and the pressure is always negative, but the sum $\rho+3p$ change its sign from positive at early times to negative at intermediate times and again to positive at later times, so that we have three distinct periods for the acceleration field: accelerated-decelerated-accelerated.

\begin{figure}[h]\vskip0.8cm
\centerline{\includegraphics[width=8.5cm]{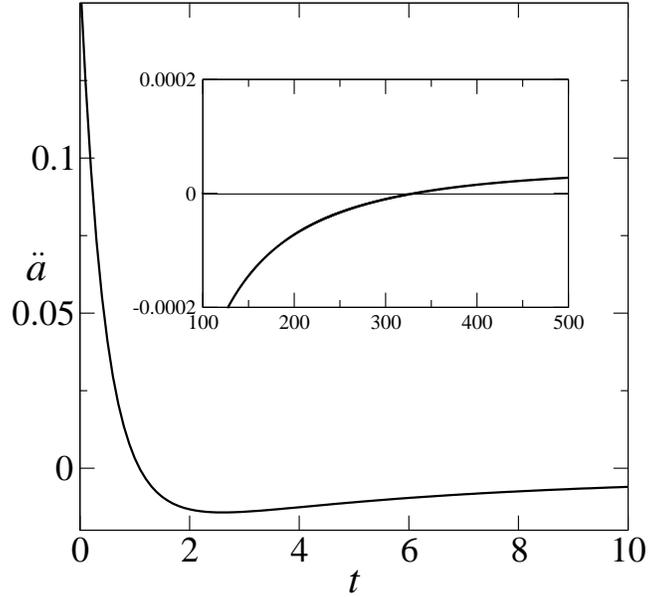}}
\caption{ Acceleration field $\ddot a$ versus time $t$.}\end{figure}

\section{Conclusions}

In this work we have analyzed a cosmological model with two constituents, a tachyonic and a fermionic field interacting via a Yukawa-type potential type. To sum up: the tachyonic field
answers for the initial accelerated regime and for the subsequent decelerated regime so that it behaves as an inflaton at early times and as a matter field at intermediate times, while the fermionic field has the role of a dark energy constituent, since it leads to an accelerated regime at later times. The interaction between both fields  occurs via a Yukawa-type potential
 which controls the extension in time of the
decelerated period of this universe: in fact,  a stronger coupling implies
into a shorter  matter dominated era.
\section*{Acknowledgments}

One of the authors (GMK) acknowledges the financial support of Conselho Nacional de Desenvolvimento
Cient\'{\i}fico e Tecnol\'ogico -- CNPq  (Brazil).

\end{document}